\documentclass[twocolumn,aps,prb,superscriptaddress,groupedaddress]{revtex4}
\usepackage{dcolumn}% Align table columns on decimal point
\usepackage{bm}% bold math
\usepackage{amsmath,amssymb}
\usepackage[dvipdfmx]{graphicx}
\usepackage{color}
\usepackage{soul}
\sethlcolor{yellow}
\usepackage{braket}
\usepackage{appendix}

\begin{document}

\title{Tunable tunnel coupling in a double quantum antidot with cotunneling
via localized state}
\author{Tokuro Hata}
\affiliation{Department of Physics, Tokyo Institute of Technology, 2-12-1 Ookayama,
Meguro, Tokyo 152-8551, Japan.}
\author{Kazuhiro Sada}
\affiliation{Department of Physics, Tokyo Institute of Technology, 2-12-1 Ookayama,
Meguro, Tokyo 152-8551, Japan.}
\author{Tomoki Uchino}
\affiliation{Department of Physics, Tokyo Institute of Technology, 2-12-1 Ookayama,
Meguro, Tokyo 152-8551, Japan.}
\author{Daisuke Endo}
\affiliation{Department of Physics, Tokyo Institute of Technology, 2-12-1 Ookayama,
Meguro, Tokyo 152-8551, Japan.}
\author{Takafumi Akiho}
\affiliation{NTT Basic Research Laboratories, 3-1,Morinosato Wakamiya, Atsugi-shi,
Kanagawa 243-0198, Japan.}
\author{Koji Muraki}
\affiliation{NTT Basic Research Laboratories, 3-1,Morinosato Wakamiya, Atsugi-shi,
Kanagawa 243-0198, Japan.}
\author{Toshimasa Fujisawa}
\affiliation{Department of Physics, Tokyo Institute of Technology, 2-12-1 Ookayama,
Meguro, Tokyo 152-8551, Japan.}
\date{\today }

\begin{abstract}
Controlling tunnel coupling between quantum antidots (QADs) in the quantum
Hall (QH) regime is problematic. We propose and demonstrate a scheme for tunable tunnel coupling between two QADs by utilizing a cotunneling process via a localized
state as a third QAD. The effective tunnel coupling can be tuned by changing
the localized level even with constant nearest-neighbor tunnel couplings. We
systematically study the variation of transport characteristics in the
effectively triple QAD system at the Landau level filling factor $\nu =2$.
The tunable tunnel coupling is clarified by analyzing the anti-crossing of
Coulomb blockade peaks in the charge stability diagram, in agreement with
numerical simulations based on the master equation. The scheme is attractive
for studying coherence and interaction in QH systems.
\end{abstract}

\maketitle

\section{Introduction}

Controlling quantum coherence and interaction between particles is a central
subject in mesoscopic physics of conventional and topological states of
matter. In the case of the integer and fractional quantum Hall (QH) regimes,
the interplay between the Aharonov-Bohm effect and the Coulomb blockade (CB)
effect has been discussed for small filled regions (quantum dots, QDs)~\cite%
{Sivan2016,Roosli2020,Roosli2021} and small empty regions (quantum antidots,
QADs)~\cite{HwangPRB1991, FordPRB1994, KataokaPRL1999,
SimPhysRep2007,MillsPRB2019, MillsPRL2020}. Understanding the two effects is
essential to unveil the anyon statistics of fractional charges~\cite%
{HalperinPRB2011,BerndPRL2020,NakamuraNatPhys2019, NakamuraNatPhys2020}.
Double QDs and QADs should provide another platform for better control of
coherency and interaction. Particularly, double QADs allow us to study
coherent tunneling of quasiparticles~\cite{AverinPhysicaE2007}. However, in
contrast to the successful development of quantum information devices with
QDs~\cite{WielRevModPhys2003,HayashiPRL2003,PettaScience2005}, controlling
tunnel and electrostatic couplings in double QADs remains challenging even
in the integer quantum Hall regime. While a few papers report on the tunnel
and electrostatic coupling of double QADs, finite tunnel couplings are confirmed only at some particular conditions without tuning capability~\cite{GouldPRL1996,MaasiltaPRL2000}.
Tunable coupling strength is highly desirable for manipulating
quasiparticles. {The issue might be related to the formation of tunnel
barriers in a QH insulator with a narrow energy gap determined by the cyclotron and Zeeman energies~\cite{Ezawa2013}.
The barrier height cannot exceed the energy gap, and localized states randomly
distributed in the QH insulator are unexpectedly charged or discharged.
Smooth control of tunnel coupling may not be available with standard
techniques.}

Here, we propose a triple QAD configuration to control the tunnel coupling
between the outer two QADs by using the second-order tunneling process
through the central QAD. The basic characteristics are investigated by using
a localized state acting as the central QAD located between well-controlled
QADs with gates. The charge stability diagram shows a dramatic change from
the parallelogram pattern showing negligible coupling between the two QADs
to the rounded honeycomb pattern manifesting the presence of tunnel coupling
by controlling the energy level of the localized state. The transition is
consistent with a model calculation involving the hybridization of the
electronic states in the triple QAD. The scheme might be useful in studying
coherent tunneling of quasiparticles in a controllable way.

\begin{figure*}[tbp]
\center \includegraphics[width=\linewidth]{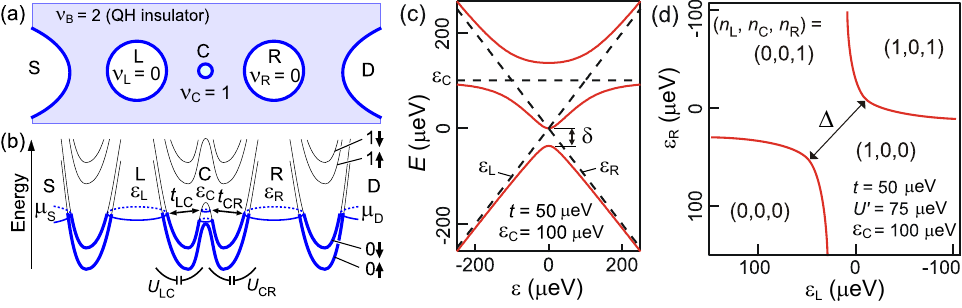}
\caption{(a) Schematic illustration of a triple QAD (L, C, and R) between
the source (S) and drain (D) in a QH insulator at $\protect\nu _{\mathrm{B}%
}=2$. Local filling factors $\protect\nu _{\mathrm{L}}=0$, $\protect\nu _{%
\mathrm{C}}=1$, and $\protect\nu _{\mathrm{R}}=0$ are assumed. (b) Energy
diagram of spin-resolved Landau levels (0$\uparrow $, 0$\downarrow $, 1$%
\uparrow $, and 1$\downarrow $) for the triple QAD. Bound states with
energies $\protect\varepsilon _{\mathrm{L}}$, $\protect\varepsilon _{\mathrm{%
C}}$, and $\protect\varepsilon _{\mathrm{R}}$ are coupled with tunnel
coupling $t_{\mathrm{LC}}$ and $t_{\mathrm{CR}}$, and electrostatic coupling 
$U_{\mathrm{LC}}$ and $U_{\mathrm{CR}}$. (c) Eigenenergies of a one-electron
triple QAD with $t=50\ \mathrm{\protect\mu eV}$ (the solid lines) and $t=0$
(the dashed lines) as a function of $\protect\varepsilon \equiv \protect%
\varepsilon _{\mathrm{L}}-\protect\varepsilon _{\mathrm{R}}$. (d) Stability
diagram of charge states $(n_{\mathrm{L}},n_{\mathrm{C}},n_{\mathrm{R}})$
for the triple QAD with $t=50\ \mathrm{\protect\mu eV}$, $U^{\prime }=75\ 
\mathrm{\protect\mu eV}$, and $\protect\varepsilon _{\rm C}=100\ \mathrm{\protect%
\mu eV}$.}
\end{figure*}

\section{Triple QAD system}

We consider a triple QAD system at Landau-level filling factor $\nu _{%
\mathrm{B}}=2$ in the bulk, as shown in Fig.~1(a). The left (L), central
(C), and right (R) QADs are formed with local filling factors, $\nu _{%
\mathrm{L}}=0$, $\nu _{\mathrm{C}}=1$, and $\nu _{\mathrm{R}}=0$,
respectively, between the source (S) and drain (D) regions, where $\nu _{%
\mathrm{C}}=1$ is assumed for a localized state as QAD C in this paper. The
following scheme should work even for other filling factors. The energy
diagram of the system is schematically shown in Fig. 1(b) for spin-up and
-down branches of the lowest and second-lowest Landau levels. Transport is
dominated by tunneling through bound states, $\varepsilon _{\mathrm{L}}$, $%
\varepsilon _{\mathrm{C}}$, and $\varepsilon _{\mathrm{R}}$, in the
spin-down lowest Landau level (0$\downarrow $). The tunnel coupling, $t_{%
\mathrm{LC}}$ and $t_{\mathrm{CR}}$, and the electrostatic coupling, $U_{%
\mathrm{LC}}$ and $U_{\mathrm{CR}}$, should be determined by the potential
profile of the QH insulator. Notice that the bulk is insulating only near
the QH filling factor ($\nu _{\mathrm{B}}=2$ in our case). Deviation from
the integer value induces occupation of integer charges on localized states
randomly distributed in the sample, which alters the potential profile.
Therefore, standard techniques, such as surface gates that change the
electron density underneath and adjusting magnetic field that changes the
flux density, may not provide smooth control of tunnel coupling. Here, we
use $\varepsilon _{\mathrm{C}}$ as a control knob to induce tunnel and
electrostatic coupling between QADs L and R by utilizing second-order
tunneling.

The hybridization of QADs L, C, and R can be described by the effective
one-electron Hamiltonian 
\begin{equation}
H_{1}=\left( 
\begin{array}{ccc}
\varepsilon _{\mathrm{L}} & t_{\mathrm{LC}} & 0 \\ 
t_{\mathrm{LC}}^{\ast } & \varepsilon _{\mathrm{C}} & t_{\mathrm{CR}} \\ 
0 & t_{\mathrm{CR}}^{\ast } & \varepsilon _{\mathrm{R}}%
\end{array}%
\right)  \label{Eq001}
\end{equation}%
for the first electron from the reference electron numbers in the system.
Here, we consider only the nearest-neighbor tunneling $t_{\mathrm{LC}}$ and $%
t_{\mathrm{CR}}$ by neglecting distant tunneling between L and R. Only a
single energy level in each QAD is considered for simplicity. Figure~1(c)
shows the eigenenergies of the system as a function of the energy bias $%
\varepsilon \equiv \varepsilon _{\mathrm{L}}-\varepsilon _{\mathrm{R}}$ for $%
\varepsilon _{\mathrm{L}}=\varepsilon /2$ and $\varepsilon _{\mathrm{R}%
}=-\varepsilon /2$ at $\varepsilon _{\mathrm{C}}=100\ \mathrm{\mu eV}$ and $%
t_{\mathrm{LC}}=t_{\mathrm{CR}}=t=50\ \mathrm{\mu eV}$ (the solid lines). As
compared to the uncoupled case with $t_{\mathrm{LC}}=t_{\mathrm{CR}}=0$ (the
dashed lines), finite energy splitting $\delta $ is seen in the hybridized
states (the solid lines) around the crossing of $\varepsilon _{\mathrm{L}}$
and $\varepsilon _{\mathrm{R}}$. The bottom trace shows the ground-state
energy $E_{1}^{(\mathrm{GS})}$ of the one-electron system. The splitting $%
\delta =\left( \sqrt{\varepsilon _{\mathrm{C}}^{2}+8t^{2}}-\left\vert
\varepsilon _{\mathrm{C}}\right\vert \right) /2$ at $\varepsilon _{\mathrm{L}%
}=\varepsilon _{\mathrm{R}}=0$ is tunable with $\varepsilon _{\mathrm{C}}$
even when $t$ is fixed. The second-order tunneling can be seen in the
approximated form of $\delta \simeq 2t^{2}/\left\vert \varepsilon _{\mathrm{C%
}}\right\vert $ for $\left\vert \varepsilon _{\mathrm{C}}\right\vert \gg t$.

We study the higher-order tunneling by investigating the charging diagram of
the triple QADs. The system accommodates ($n_{\mathrm{L}}$, $n_{\mathrm{C}}$%
, $n_{\mathrm{R}}$) numbers of excess electrons in the respective QADs by
varying $\varepsilon _{\mathrm{L}}$ and $\varepsilon _{\mathrm{R}}$. In the
absence of distant electrostatic coupling between L and R, the two-electron
Hamiltonian reads 
\begin{equation}
H_{2}=\left( 
\begin{array}{ccc}
\varepsilon _{\mathrm{L}}+\varepsilon _{\mathrm{C}}+U_{\mathrm{LC}} & t_{%
\mathrm{CR}} & 0 \\ 
t_{\mathrm{CR}}^{\ast } & \varepsilon _{\mathrm{L}}+\varepsilon _{\mathrm{R}}
& t_{\mathrm{LC}} \\ 
0 & t_{\mathrm{LC}}^{\ast } & \varepsilon _{\mathrm{R}}+\varepsilon _{%
\mathrm{C}}+U_{\mathrm{CR}}%
\end{array}%
\right)  \label{Eq002}
\end{equation}%
for the charge bases $(1,1,0)$, $(1,0,1)$, and $(0,1,1)$. The ground-state
energy $E_{2}^{(\mathrm{GS})}$ of the two-electron system can be obtained by
diagonalizing $H_{2}$. The system takes the charge state $(n_{\rm L},n_{\rm C},n_{\rm R})$ with minimum
energy, as shown in the stability diagram of Fig. 1(d) in the $\varepsilon _{%
\mathrm{L}}-\varepsilon _{\mathrm{R}}$ plane. The boundaries among three
regions with different total electron number $n=n_{\mathrm{L}}+n_{\mathrm{C}%
}+n_{\mathrm{R}}$ are shown by the red lines. Here, we investigate the minimum spacing $%
\Delta $ between the charge states $(0,0,0)$ and $(1,0,1)$ in this paper.
For $t_{\mathrm{LC}}=t_{\mathrm{CR}}\equiv t$ and $U_{\mathrm{LC}}=U_{%
\mathrm{CR}}\equiv U^{\prime }$, $\Delta $ is given by 
\begin{equation}
\Delta =\frac{U^{\prime }-\left\vert \varepsilon _{\mathrm{C}}\right\vert +2%
\sqrt{\varepsilon _{\mathrm{C}}^{2}+8t^{2}}-\sqrt{\left( \left\vert
\varepsilon _{\mathrm{C}}\right\vert +U^{\prime }\right) ^{2}+8t^{2}}}{2},
\label{SpacingExact}
\end{equation}%
which includes $\delta =\frac{1}{2}\left( \sqrt{\varepsilon _{\mathrm{C}%
}^{2}+8t^{2}}-|\varepsilon _{\mathrm{C}}|\right) $. The remainder $U^{\prime
\prime }\equiv \Delta -\delta $ can be understood as emergent electrostatic
coupling induced by the second-order tunneling ($U^{\prime \prime }\simeq
2t^{2}\frac{U^{\prime }}{|\varepsilon _{\mathrm{C}}|\left( |\varepsilon _{%
\mathrm{C}}|+U^{\prime }\right) }$ for $\left\vert \varepsilon _{\mathrm{C}%
}\right\vert \gg t$). Therefore, observation of finite $\Delta $ induced at
small $\left\vert \varepsilon _{\mathrm{C}}\right\vert $ suggests tunable
coupling of $\delta $ and $U^{\prime \prime }$. Note that symmetric
parameters ($t_{\mathrm{LC}}=t_{\mathrm{CR}}$ and $U_{\mathrm{LC}}=U_{%
\mathrm{CR}}$) are assumed for simplicity, and tunable coupling is expected
even with asymmetric parameters.

{The model is equivalent to that for triple QDs. }While similar three-level
systems can be seen in previous studies on QDs and atoms~\cite%
{TakakuraAPL2014,WaughPRL1995, GroveRasmussenNanoLett2008}, their
realization in QADs would provide a significant step for coherent control of
quasiparticles.

\begin{figure}[tbp]
\center \includegraphics[width=\linewidth]{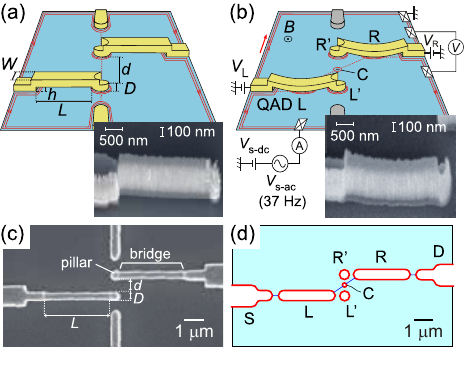}
\caption{(a) Schematic illustration of the original QAD design with two
airbridge gates and two side gates. Two QADs are expected to form around the
pillars. The inset shows the $45^{\circ }$-tilt SEM image of an undeformed
bridge used in our previous experiments~\protect\cite%
{EguchiAPEX2019,HataJJAP2020}. (b) Schematic illustration of the present
sample with two deformed gates. Measurement setup for transport through QAD
L and R under the deformed bridge and impurity-induced QAD C is shown. The
inset shows the $45^{\circ }$-tilt SEM image of the deformed airbridge
(shrunk horizontally to highlight the deformation). {(c) A top-view SEM
image of the present sample with the deformed bridges (taken after the
measurement). (d) Schematic geometry of QAD L, C, and R between the source
(S) and drain (D) channels. The QAD L' and R' might be absorbed in L and R.}}
\end{figure}

\section{Experiment}

\subsection{Sample and measurement setup}

Our sample was fabricated in a standard AlGaAs/GaAs heterostructure with
two-dimensional electron gas located at 100 nm below the surface. With an
electron density of $\sim 2.75\times 10^{15}\ \mathrm{m^{-2}}$, a QH state
at $\nu _{\mathrm{B}}=2$ can be prepared by applying perpendicular magnetic
field $B\simeq 5\ \mathrm{T}$. Two airbridge gates with Ti (thickness of $%
30\ \mathrm{nm}$) and Au ($270\ \mathrm{nm}$) layers were fabricated by
using electron-beam lithography with a triple layer resist \cite%
{EguchiAPEX2019,HataJJAP2020}. Each gate has a small pillar of diameter $%
D=300\ \mathrm{nm}$ and is connected to the lead electrode through the
bridge of length $L=3\ \mathrm{\mu m}$, width $W=300\ \mathrm{nm}$, and
bridge height $h=150\ \mathrm{nm}$, as shown in Fig.~2(a). The two pillars
are separated by distance $d=500\ \mathrm{nm}$. This device was originally
designed to form two QADs around the pillars [the red circles in Fig.~2(a)].
Such airbridge gates worked nicely in our previous paper~\cite%
{EguchiAPEX2019,HataJJAP2020}. However, for the particular device used in
this work, it turned out that the airbridges have been deformed, as shown in
Fig.~2(b) with an SEM picture taken after the measurement. The central part
of the bridge is touching the surface of the heterostructure. We noticed
later that the deformation was introduced during the post photo-lithography
process with PMGI, which was not used for the previous devices. Note that
the deformation of the bridge is reproducible with the same process, whereas
the detailed mechanism of the deformation is not known.

As a result of the deformation, a relatively large QAD with the area of $%
LW\simeq 1\ (\mathrm{\mu m})^{2}$ should be formed under the deformed
bridge. This area is comparable to those of typical QADs seen in the
literature~\cite{KouPRL2012}. We find that such QADs, referred to as QADs L
and R, under the deformed bridges work nicely in this work. However, {we did
not find any characteristics associated with the intended QADs L' and R'
under the pillars [see Section III-C]. }

We take advantage of {localized} states present in our device. While they
are randomly distributed in the sample, we {focus on a specific localized
state, which acts as QAD C, located between QADs L and R. Following
measurements suggests that QAD\ C with the area of $0.02-0.04\ \mathrm{(\mu m)^{2}%
}$, equivalent to a circle with a diameter of $160-230\ \mathrm{nm}$}, is
located in the middle of QAD L and R{, as illustrated in Fig.~2(b) [see
Section III-B]. Figure~2(c) shows {the} top view SEM image of the present
device taken after the measurements, where pillars, (deformed) airbridges,
and lead electrodes are seen. Schematic locations of QADs are illustrated
in Fig.~2(d), while QADs L' and R' might be merged into L and R,
respectively. }

The transport through the QADs is investigated by applying AC voltage $V_{%
\mathrm{s-ac}}=30\ \mathrm{\mu V}$ at $37\ \mathrm{Hz}$ and DC voltage $V_{%
\mathrm{s-dc}}$ (= 0 unless otherwise noted) to the source and measuring the
AC voltage drop $V$ between the voltage probes with a lock-in amplifier
[Fig. 2(b)]. The differential conductance $G$ is estimated from the relation 
$G=(2e^{2}/h)\times (V/V_{\mathrm{s-ac}})$. All measurements were performed
in a dilution refrigerator with a base temperature of about $100\ \mathrm{mK}
$.

\begin{figure}[tbp]
\includegraphics[width=\linewidth]{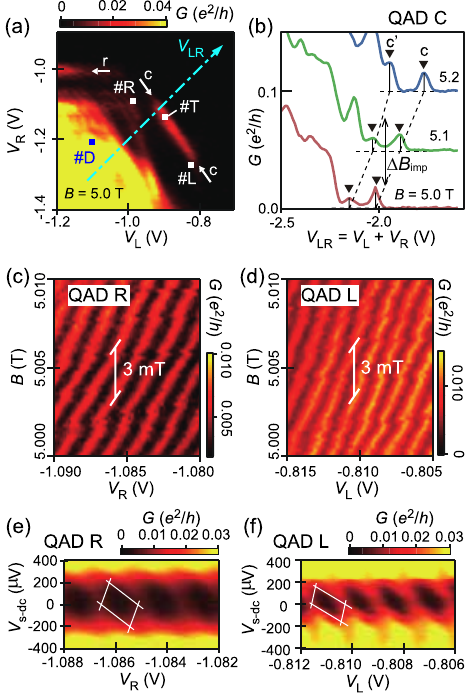}
\caption{(a) Color plot of differential conductance $G$ as a function of $V_{%
\mathrm{L}}$ and $V_{\mathrm{R}}$ at $B=5.0\ \mathrm{T}$ and $V_{\mathrm{s-dc%
}}$ = 0. The peak between the arrows labeled c is the CB peak associated
with QAD C. Single QADs L and R are investigated at conditions marked by \#L
and \#R, respectively. Double and triple QADs are investigated at conditions
marked by \#D and \#T, respectively. (b) $G$ as a function of $V_{\mathrm{LR}%
}=V_{\mathrm{L}}+V_{\mathrm{R}}$ along the dot-dashed line in (a) at
different magnetic fields. Each trace is offset by $0.05e^{2}/h$ for
clarity. (c) and (d) $G$ as a function of $B$ and each gate voltage. (e) and
(f) Coulomb diamond characteristics seen in $G$ as a function of bias
voltage $V_{\mathrm{s-dc}}$ and the gate voltage, $V_{\mathrm{R}}$ for QAD R
in (e) and $V_{\mathrm{L}}$ for QAD L in (f). The white parallelogram shows
an approximate CB region.}
\end{figure}

\subsection{Localized state as QAD C}

Figure 3(a) shows a color plot of conductance $G$ over the wide ranges of $%
V_{\mathrm{L}}$ and $V_{\mathrm{R}}$ at $B=5.0\ \mathrm{T}$. Whereas the
Coulomb oscillations of QAD L and R are not visible with this coarse scan,
several current peaks associated with localized states are resolved. For
example, the horizontal line at $V_{\mathrm{R}}\simeq -1.0$ V (marked by the
arrow labeled r)\ should be associated with a localized state near the right
gate with $V_{\mathrm{R}}$ but far from the left gate with $V_{\mathrm{L}}$.
We focus on a specific localized state that exhibits the current peak marked
by the arrows labeled c. As this peak is elongated in the lower-right
direction with a slope of $dV_{\mathrm{R}}/dV_{\mathrm{L}}=-1.45$ in the
figure, the state is almost equally coupled to the two gates, and thus
should be located at around the center of the two gates (slightly closer to
the left gate). We shall use this localized state as QAD C in the following.

Figure~3(b) shows the conductance traces taken by simultaneously changing $%
V_{\mathrm{L}}$ and $V_{\mathrm{R}}$ along the dot--dashed line labeled $V_{%
\mathrm{LR}}$ in Fig.~3(a) for several $B$ values. In Fig. 3(b), the two
peaks labeled c and c' evolve in a similar manner with $B$, as shown by the
dashed lines, implying that they are two consecutive CB peaks for the same impurity.
The corresponding magnetic-field period of about $\Delta B_{\mathrm{imp}}=0.1-0.2\ {\rm T}$ suggests that the area enclosed by the bound state is $S_{\mathrm{imp}}=(h/e)/\Delta B_{\mathrm{imp}}\simeq $ $0.02-0.04\ \mathrm{(\mu m)^{2}}$ by assuming local filling factor $\nu _{\mathrm{C}}=1$. If the
bound state is circular, its diameter of $160-230\ \mathrm{nm}$ can fit in
between the two gates with the distance of $d=$ 500 nm, as illustrated in
Fig.~2(d).

In Section II-E, we focus on peak c of Fig. 3(b), where rich characteristics
associated with QADs L, C, and R show up in the fine sweep of $V_{\mathrm{R}%
} $ and $V_{\mathrm{L}}$. 
\begin{figure}[tbp]
\includegraphics[width=\linewidth]{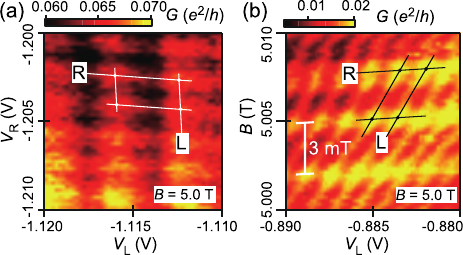}
\caption{(a) Color plot of conductance $G$ as a function of $V_{\mathrm{L}}$
and $V_{\mathrm{R}}$ at around condition \#D in Fig. 3(a) for uncoupled QADs
L and R showing parallelogram patterns (the white lines). (b) $G$ as a
function of $V_{\mathrm{L}}$ and $B$ at $V_{\mathrm{R}}=$ $-1.135\ {\rm V}$, showing
another example of uncoupled QADs L and R. The black lines labeled L and R
represent CB peaks attributed to QAD L and R, respectively. }
\end{figure}

\subsection{Single QAD L and R}

The QADs L and R were investigated separately by focusing on the conditions
\#L and \#R, respectively, in the $V_{\mathrm{R}}-V_{\mathrm{L}}$\ plane of
Fig. 3(a). The transport is effectively determined by each QAD under the
asymmetric gate voltages, where other QADs are strongly coupled to the leads.
CB oscillations of QAD R can be seen in fine
sweeps of $V_{\mathrm{R}}$ and $B$, as shown in Fig.~3(c). Its oscillation
period in $B$ is $\Delta B_{\mathrm{R}}\simeq $ 3 mT, which corresponds to
the area enclosed by the bound states, $S_{\mathrm{R}}=\frac{1}{2}\frac{h}{e}%
\frac{1}{\Delta B_{\mathrm{R}}}\simeq $ 0.7 ($\mu $m)$^{2}$. Here, the
factor $\frac{1}{2}$ is used for the two occupied spin-resolved Landau
levels with $\nu _{\mathrm{B}}-\nu _{\mathrm{R}}=2$, where bound states
associated with the two Landau levels strongly interacted~electrostatically \cite{HwangPRB1991, FordPRB1994,
KataokaPRL1999, SimPhysRep2007}. This $S_{\mathrm{R}}$ is comparable to the
area of the deformed bridge [$LW=1.0\ \mathrm{(\mu m)^{2}}$] but far from
the area of the pillar [$\sim ${$0.07\ \mathrm{(\mu m)^{2}}$})] in
consistency with QAD R being formed under the deformed gate.

\begin{figure*}[tbp]
\center \includegraphics[width=170mm]{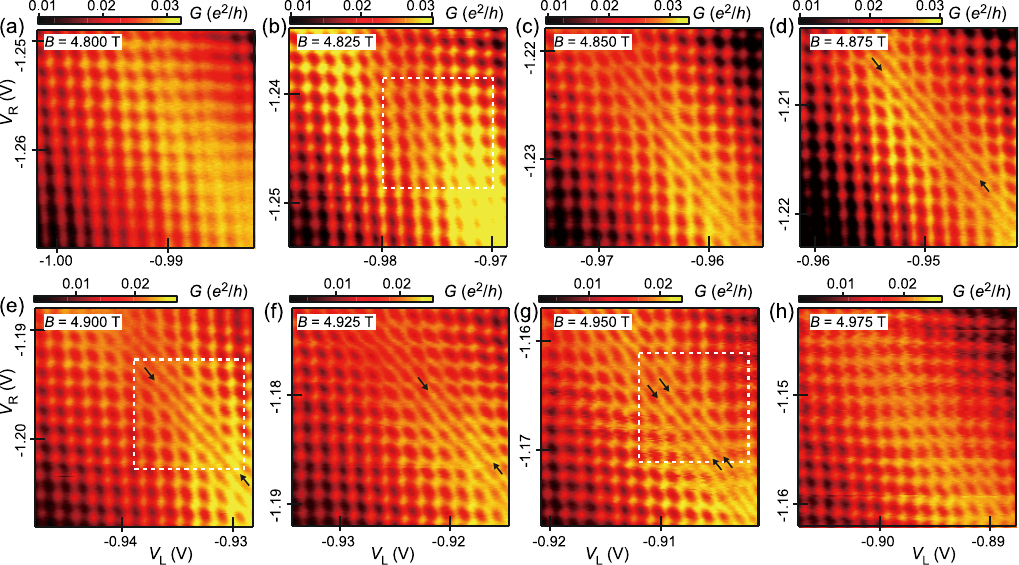}
\caption{(a)--(h) Color plots of $G$ as a function of $V_{\mathrm{L}}$ and $%
V_{\mathrm{R}}$ at different magnetic fields from $4.800\ \mathrm{T}$ in (a)
to $4.975\ \mathrm{T}$ in (h). For each $B$, the sweep ranges of $V_{\mathrm{%
L}}$ and $V_{\mathrm{R}}$ are adjusted to focus on the broad CB peak of the
impurity QAD C. Some diagonal straight lines are highlighted by the arrows.}
\end{figure*}

The Coulomb diamond characteristics of QAD R are obtained by applying $V_{%
\mathrm{s-dc}}$ and $V_{\mathrm{R}}$, as shown in Fig.~3(e). The CB region
with $G\sim 0$ is seen in the voltage range $\left\vert V_{\mathrm{s-dc}%
}\right\vert \lesssim $ 200 {$\mathrm{\mu V}$, as illustrated by a white }%
parallelogram as a guide. This measures the addition energy $U_{\mathrm{R}%
}\simeq $ 200 {$\mathrm{\mu eV}$}, which includes the on-site Coulomb
charging energy and level spacing of the bound states. This value is
comparable to {typical values of reported QADs with similar sizes~\cite%
{KouPRL2012}. The energy of }each bound state can be shifted by $\alpha _{%
\mathrm{R}}\Delta V_{\mathrm{R}}$ with small change $\Delta V_{\mathrm{R}}$
in $V_{\mathrm{R}}$, where the lever arm factor $\alpha _{\mathrm{R}}\simeq
-0.13e$ is roughly estimated from the size of the parallelogram.

Similarly, QAD L investigated at around condition \#L shows CB oscillations
in Fig.~3(d). The oscillation period $\Delta B_{\mathrm{L}}\simeq $ 3 mT
also suggests that the QAD is formed under the deformed bridge. The two QADs
show similar oscillation periods in $B$ as well as their gate voltages ($V_{%
\mathrm{R}}$ and $V_{\mathrm{L}}$). The Coulomb diamond characteristics for
QAD L in Fig. 3(f) show smaller blockade regions with somewhat smaller
addition energy $U_{\mathrm{L}}\simeq $ 180 {$\mathrm{\mu eV}$} and $\alpha
_{\mathrm{L}}\simeq -0.15e$. Similar QADs with small differences are well reproduced by the deformed bridges.

\subsection{Uncoupled double QAD}

Transport through QADs L and R with a negligible role of QAD C can be seen
when large negative voltages, $V_{\mathrm{L}}$ and $V_{\mathrm{R}}$, are
applied. Figure 4(a) shows such Coulomb oscillations in the fine sweep of $%
V_{\mathrm{L}}$ and $V_{\mathrm{R}}$ at around condition \#D in Fig. 3(a).
The oscillations for QAD L (the vertical lines) and R (the horizontal lines)
are resolved but not influenced by each other with no measurable splitting
at their crossings, as shown by the white parallelogram in Fig. 4(a). This
is the signature of negligible tunnel and electrostatic couplings, as
studied with conventional QDs~\cite{WielRevModPhys2003}.

Another example of uncoupled QADs is shown in the $B-V_{\mathrm{L}}$ plane
of Fig. 4(b), where the parallelogram pattern (the black lines) for CB
oscillations of QAD L and R is resolved.
Whereas this $V_{\mathrm{L}}$ and $%
V_{\mathrm{R}}$ range is the condition where triple QAD formation is
expected [\#T in Fig. 3(a)], a negligible role of QAD C is seen probably due to small tunneling ($t_{\rm LC}$, $t_{\rm RC}$) in this $B$
range. The magnetic field periods, $\Delta B_{\mathrm{R}}\simeq $ 3 mT and $%
\Delta B_{\mathrm{L}}\simeq $ 3 mT, are similar to those obtained for single
QADs. Therefore, QADs L and R are stably formed in the wide range of $V_{%
\mathrm{L}}$ and $V_{\mathrm{R}}$.

The above data show that distant QADs L and R are uncoupled with negligible
tunneling ($t_{\mathrm{LR}}\simeq 0$) and electrostatic ($U_{\mathrm{LR}%
}\simeq 0$) couplings. However, the two QADs can be coupled by introducing
QAD C, as shown in the next subsection.

\begin{figure*}[tbp]
\includegraphics[width=\linewidth]{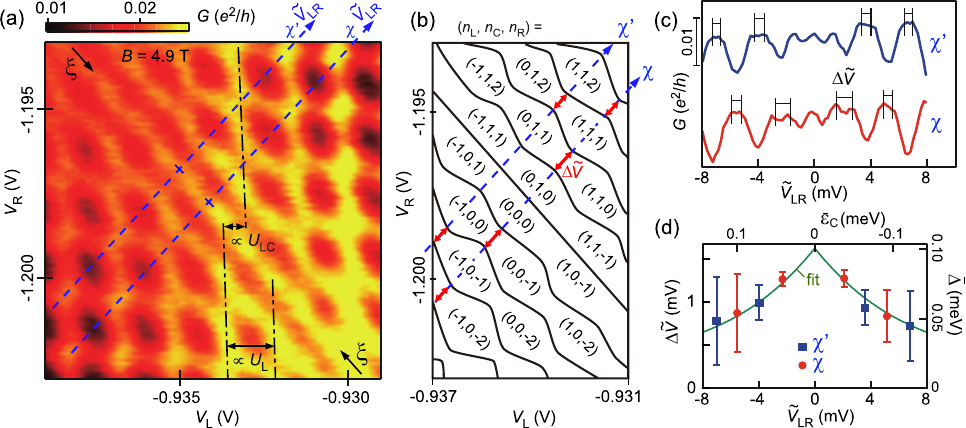}
\caption{(a) $G$ as a function of $V_{\mathrm{L}}$ and $V_{\mathrm{R}}$
measured at $4.900\ \mathrm{T}$ [equivalent gate-voltage condition is marked
by \#T in Fig. 3(a)]. The straight diagonal line is indicated by two arrows
labeled $\protect\xi $. Other honeycomb and parallelogram patterns are
almost symmetric about the diagonal line. The spacings of CB oscillations
(the dot-dashed lines) for QAD L reflect $U_{\mathrm{L}}$ and $U_{\mathrm{LC}%
}$. (b) Schematic charge stability diagram with excess electron numbers $%
\left( n_{\mathrm{L}},n_{\mathrm{C}},n_{\mathrm{R}}\right) $ for (a). The
red arrows indicate the spacing $\Delta \tilde{V}$ of the anticrossings. (c)
Cross-sectional slices of $G$ along the lines $\protect\chi $ and $\protect%
\chi ^{\prime }$ as a function of relative gate voltage $\tilde{V}_{\mathrm{%
LR}}$ in $V_{\mathrm{LR}}=V_{\mathrm{L}}+V_{\mathrm{R}}$ measured from the
straight diagonal line in (a). The anti-crossing is represented by the vertical lines as guides to the eye.  (d) The anticrossing $\Delta \tilde{V}$ in $%
V_{\mathrm{LR}}$ is plotted as a function of $\tilde{V}_{\mathrm{LR}}$ for
the slices $\protect\chi $ and $\protect\chi ^{\prime }$ in (c). The right
axis $\bar{\Delta}=\left\vert \frac{1}{2}\protect\alpha _{\Delta }\Delta 
\tilde{V}\right\vert $ is converted with $\protect\alpha _{\Delta }=-0.13e$.
The top axis $\bar{\protect\varepsilon}_{\mathrm{C}}$ is estimated from the
fit (the green line) with the model.}
\end{figure*}

\subsection{Triple QAD}

Coupling of QADs L, C, and R can be found in Fig.~5, where $G$ was measured
with a fine sweep of $V_{\mathrm{L}}$ and $V_{\mathrm{R}}$ at various
magnetic fields. The sweep ranges of $V_{\mathrm{L}}$ and $V_{\mathrm{R}}$
are adjusted for each $B$ to keep the focus on the resonance with QAD C [the
equivalent condition is marked by \#T in Fig.~3(a)]. Fine oscillations are
superimposed on the broad peak of QAD C. Rich characteristics ranging from
parallelogram to honeycomb patterns are seen. In addition, sharp diagonal
lines (some marked by the arrows) show up in the limited range of 4.875 T $%
\leq B\leq $ 4.950 T. Some representative characteristics are analyzed in
the following.

First, we focus on the white dashed region of Fig. 5(e) at $B=$ 4.900 T,
which is enlarged in Fig. 6(a). Strikingly, one of the CB peaks follows the
single straight diagonal line between the two arrows labeled $\xi $. This
line is located around the center of the CB peak of QAD C, and the line
width is much sharper than the peak width of QAD C. The meaning of the
diagonal line is clarified by analyzing the Coulomb oscillations of QADs L
and R. The CB peaks of QAD L [the dot-dashed lines in Fig. 6(a)] are
abruptly shifted across the diagonal line. This shift suggests that the
charge state of QAD C is changed from $n_{\mathrm{C}}=0$ (the lower-left
side) to $1$ (the upper-right side), which influences the potential of QAD
L. Therefore, we define $\varepsilon _{\mathrm{C}}=0$ on the diagonal line.
The shift measures the electrostatic coupling $U_{\mathrm{LC}}\simeq 75\ 
\mathrm{\mu eV}$ between QAD L and C. Similar shift is seen for CB peaks of
QAD R (not marked), from which $U_{\mathrm{CR}}\simeq 75\ \mathrm{\mu eV}$
is estimated. It should be noted that the straight diagonal line is
associated with a special resonance of hybridized states under symmetric
conditions of the triple QAD, as elaborated in Section IV.

The current profile in the vicinity of the diagonal line shows a clear
rounded honeycomb pattern, which manifests finite coupling between QAD L and
R. The honeycomb pattern gradually changes to the parallelogram pattern
toward the upper-right and lower-left corners. The corresponding charge
stability diagram is sketched in Fig.~6(b), where each region is labeled
with excess electron numbers $\left( n_{\mathrm{L}},n_{\mathrm{C}},n_{%
\mathrm{R}}\right) $ from a reference. We investigate the minimum spacing $%
\Delta \tilde{V}$ between the rounded charge boundaries, which corresponds
to $\Delta $ in Eq.~(\ref{SpacingExact}). The overall conductance profile is
mirror symmetric about the diagonal line and periodic along the diagonal
line (the upper-left direction). This feature suggests that the spacing $%
\Delta \tilde{V}$ is dominantly changed only by $|\varepsilon _{\mathrm{C}}|$%
, i.e. the distance\ from the diagonal line. Other parameters, specifically $%
t_{\mathrm{LC}}$ and $t_{\mathrm{CR}}$,\ are unchanged within the sweep
range of $V_{\mathrm{L}}$ and $V_{\mathrm{R}}$. Otherwise, the conductance
pattern should change in a non-symmetric way. These characteristics support
the demonstration of tunable coupling with cotunneling.

Cross-sectional current profiles passing through several anticrossing
conditions are shown in Fig.~6(c). Here, two cross sections $\chi $ and $\chi
^{\prime }$ pick up different anticrossings as illustrated by the dashed
lines in Figs.~6(a) and (b). The axis $\tilde{V}_{\mathrm{LR}}$ denotes the
relative gate voltage in $V_{\mathrm{LR}}=V_{\mathrm{L}}+V_{\mathrm{R}}$
measured from the central diagonal line ($\varepsilon _{\mathrm{C}}=0$). The splitting $\Delta \tilde{V}$ is shown by the bars in Fig. 6(c).
The precise values and their errors are determined from the overall pattern in Fig. 6(a). For example, the peak (spot) slightly
elongated to the upper-right direction suggests finite splitting, even if
the two split peaks are unresolved in the cross-sectional plot. The
estimated $\Delta \tilde{V}$ is plotted as a function of $\tilde{V}_{\mathrm{%
LR}}$ in Fig.~6(d), in which the symmetric variation of $\Delta \tilde{V}$
is seen.

$\Delta \tilde{V}$ can be converted into the splitting energy, $\bar{\Delta}%
=\left\vert \frac{1}{2}\alpha _{\Delta }\Delta \tilde{V}\right\vert $, by
using the lever arm factor $\alpha _{\Delta }=-0.13e$ [See Appendix A], as
shown in the right scale. To see the consistency with the proposed scheme,
we assume equal tunnel coupling ($t_{\mathrm{LC}}=t_{\mathrm{CR}}=t$), which
will be justified in Section IV-A, and linear dependence of $\bar{\varepsilon%
}_{\mathrm{C}}=\eta \tilde{V}_{\mathrm{LR}}$ on $\tilde{V}_{\mathrm{LR}}$
with unknown factor $\eta $. Here, symbols with a bar ($\bar{\Delta}$ and $%
\bar{\varepsilon}_{\mathrm{C}}$) denote the quantities obtained for
different charge states, while the original $\Delta $ and $\varepsilon _{%
\mathrm{C}}$ are defined for a given charge state. We apply Eq.~(\ref%
{SpacingExact}) by replacing $\Delta $ and $\varepsilon _{\mathrm{C}}$ with $%
\bar{\Delta}$ and $\bar{\varepsilon}_{\mathrm{C}}$ for the fitting to the
data. By using $U'=U_{\mathrm{LC}}=U_{\mathrm{CR}}=75\ \mathrm{\mu eV}$, the
measured $\bar{\Delta}$ is well reproduced by the fitting [the solid green
line in Fig. 6(d)] with adjusted parameters, $t=50\ \mathrm{\mu eV}$ and $%
\eta =-0.02e$. Here, tuning of $\bar{\varepsilon}_{\mathrm{C}}$ is induced
by the purely capacitive effect with $\varepsilon _{\mathrm{C}}=\eta
^{\prime }\tilde{V}$ but partially compensated by the excess charge of the
QADs ($n_{\mathrm{L}}$ and $n_{\mathrm{R}}$). They are related by $\bar{%
\varepsilon}_{\mathrm{C}}=\eta ^{\prime }\tilde{V}+U_{\mathrm{LC}}n_{\mathrm{%
L}}+U_{\mathrm{CR}}n_{\mathrm{R}}$, where $n_{\mathrm{L}}=n_{\mathrm{R}}=0$
is defined for the region at $\varepsilon _{\mathrm{C}}=0$ [see Fig. 6(b)].
We obtained $\eta ^{\prime }=-0.06e$ from the relations. This $\eta ^{\prime
}$ does not contradict the realistic lever-arm factors in our sample [See
Appendix A], and supports our scheme. Therefore, the result indicates that
the total coupling energy $\Delta $, as well as the tunnel coupling $\delta $%
, are successfully controlled with the energy of the localized state in the
range of $0<\delta <\sqrt{2}t$ ($\simeq 70\ \mathrm{\mu eV}$ in the present
case).

\begin{figure}[tbp]
\center \includegraphics[width=\linewidth]{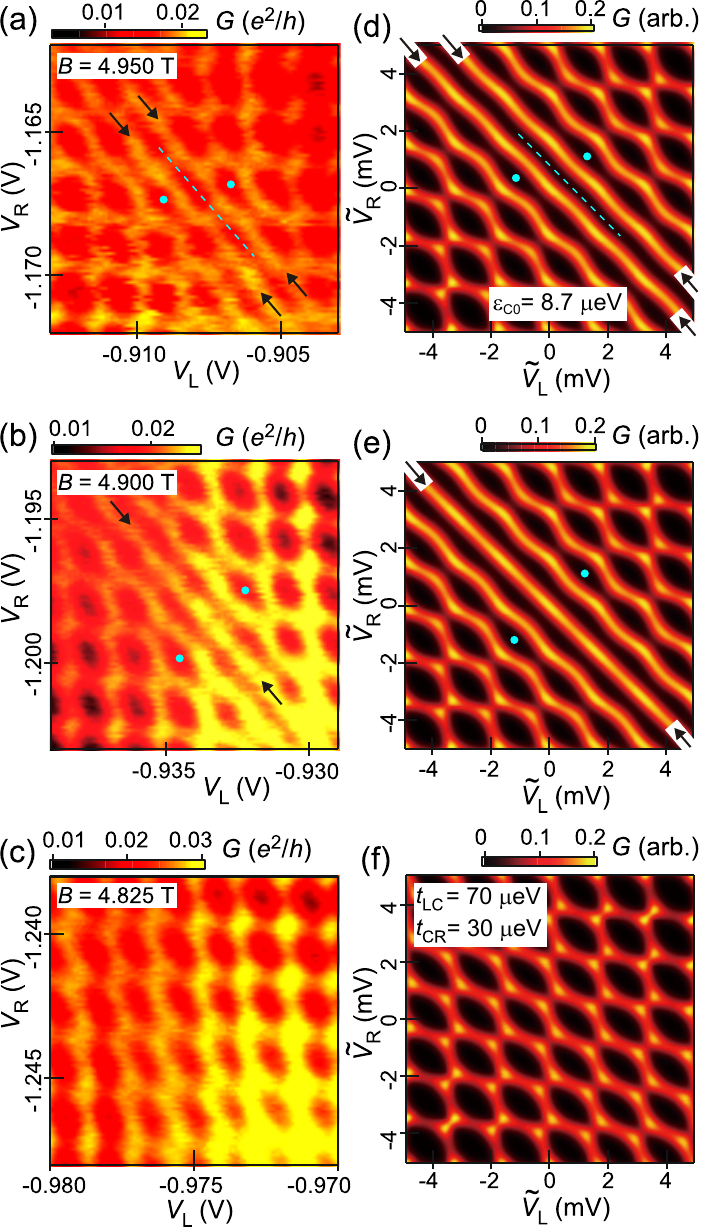}
\caption{(a-c) Measured $G$ as a function of $V_{\mathrm{L}}$ and $V_{%
\mathrm{R}}$ at $B=4.950$ T in (a), $4.900$ T in (b), and $4.825$ T in (c).
Straight diagonal lines are marked by the arrows in (a) and (b), but not
seen in (c). (d-f) Calculated $G$ as a function of $\tilde{V}_{\mathrm{L}}$
and $\tilde{V}_{\mathrm{R}}$. The offset energy for $\protect\varepsilon _{%
\mathrm{C}}$ is $\protect\varepsilon _{\mathrm{C0}}=8.7$ $\protect\mu $eV in
(d) and $\protect\varepsilon _{\mathrm{C0}}=0$ in (e) and (f). The tunnel
couplings are $t_{\mathrm{LC}}=t_{\mathrm{CR}}=50\ \mathrm{\protect\mu eV}$
in (d) and (e), and $t_{\mathrm{LC}}=70\ \mathrm{\protect\mu eV}$ and $t_{%
\mathrm{CR}}=30\ \mathrm{\protect\mu eV}$ in (f). $U_{\mathrm{L}}=U_{\mathrm{%
R}}=200\ \mathrm{\protect\mu eV}$ and $U_{\mathrm{LC}}=U_{\mathrm{CR}}=75\ 
\mathrm{\protect\mu eV}$ were used for the calculation.}
\end{figure}

The tunable tunnel coupling can be confirmed in the wide range of $B$, as
shown in Fig. 5. In all cases, one can see honeycomb patterns near the
center of the CB peak of QAD C and parallelogram patterns near the
upper-right and lower-left corners. However, the precise patterns including
the diagonal lines near the CB peak of QAD C change significantly with $B$.
The white dashed regions in Figs. 5(b), (e), and (g) are magnified in Figs.
7(b), (c), and (a), respectively. Two diagonal lines (indicated by the
arrows) are resolved at $B=4.95\ \mathrm{T}$ in Fig. 7(a), whereas only a
single line is seen at $B=4.90\ \mathrm{T}$ in Fig. 7(b) [the same data as
Fig. 6(a)].
No clear diagonal line is resolved at $B=4.825\ \mathrm{T}$ in Fig. 7(c) [see also $B=4.800$ and $4.975\ \mathrm{T}$ in Figs. 5(a) and (h), respectively], while the honeycomb pattern with finite splitting is seen.
Full understanding requires detailed analysis on the symmetry of the QAD parameters, as shown in the next section.

\section{Symmetry of triple QAD}

\subsection{System Hamiltonian}

First of all, we should note the differences in electronic states between
QADs and standard QDs. For QDs at zero or low magnetic fields, the energy levels are strongly influenced by
many-body effects with direct and exchange interactions as well as
single-particle orbitals and Zeeman splittings. As a result, CB oscillations
are generally aperiodic and thus charge stability diagrams of multiple QDs
are complicated with many jumps in the CB conditions~\cite%
{TakakuraAPL2014,WaughPRL1995,GaudreauPRL2006,SchroerPRB2007,GroveRasmussenNanoLett2008,TakakuraAPL2010,BuslNatNano2013,BraakmanNatNano2013,NoiriPRB2017,WangNanoLetters2017}
In contrast, the energy quantization of QADs in the integer QH system is
dominated by the Aharonov-Bohm effect, which determines the area of the
bound state under uniform $B$. Therefore, the energy spacing is almost
constant for a smooth QAD potential for each spin-resolved Landau level.
When multiple Landau levels are involved for each QAD, occupation of the
inner Landau level is well screened by the outer one. Therefore, CB
oscillations are periodic with a constant addition energy $U$ for different
charge states~\cite{KataokaPRB2000}.
The deviation from the periodic pattern
is studied with the hybridization of the system, as shown below.

In this paper, a symmetric QAD with $t_{\mathrm{LC}}=t_{\mathrm{CR}}$ and $U_{%
\mathrm{LC}}=U_{\mathrm{CR}}$ is assumed for proposing the tunable
second-order coupling scheme. While the symmetry is not required just for
tuning the tunnel coupling, we should investigate the role of the symmetry in the hybridization.
Interestingly, we found that the diagonal straight line observed in Fig.
6(a) is the signature of the symmetry.

Generally, CB peaks appear when the ground-state energy $E_{n}^{\left( 
\mathrm{GS}\right) }$ of the $n$-electron system coincides with that $%
E_{n+1}^{\left( \mathrm{GS}\right) }$ of the $\left( n+1\right) $-electron
system, where the electrochemical potential $\mu =E_{n+1}^{\left( \mathrm{GS}%
\right) }-E_{n}^{\left( \mathrm{GS}\right) }$ of the system equals the
chemical potentials (defined to be zero) of the leads. The appearance of the
diagonal straight line suggests that this equality ($E_{n+1}^{\left( \mathrm{%
GS}\right) }=E_{n}^{\left( \mathrm{GS}\right) }$) is satisfied on the
diagonal line over several charge states. In the presence of significant
nearest-neighbor tunneling ($t_{\mathrm{LC}},t_{\mathrm{CR}}\neq 0$), the
most probable situation is that the $n$- and $\left( n+1\right) $-electron
systems share the identical eigenenergies including the ground-state one
with the same form of Hamiltonians $H_{n}$ and $H_{n+1}$.

\begin{figure}[tbp]
\center \includegraphics[width=\linewidth]{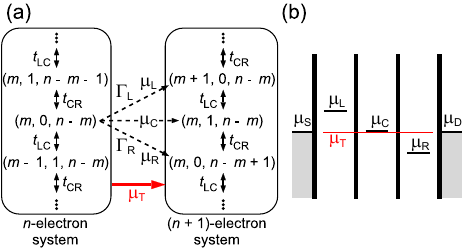}
\caption{(a) Tunnel coupling ($t_{\mathrm{LC}}$ and $t_{\mathrm{CR}}$)
between charge states $\left( n_{\mathrm{L}},n_{\mathrm{C}},n_{\mathrm{R}%
}\right) $ in $n$- and $\left( n+1\right) $-electron systems. (b) Chemical
potential diagram for the triple QAD. Sequential tunneling between the
unhybridized states can be characterized by the simple electrochemical
potentials $\protect\mu _{\mathrm{L}}$, $\protect\mu _{\mathrm{C}}$, and $%
\protect\mu _{\mathrm{R}}$. Correct electrochemical potential $\protect\mu _{%
\mathrm{T}}$ should be considered for the hybridized triple QAD.}
\end{figure}

To see this happens, the excess charges ($n_{\mathrm{L}}$, $n_{\mathrm{C}}$, 
$n_{\mathrm{R}}$) that belong to the $n$- and $\left( n+1\right) $-electron
systems are listed in Fig. 8(a) with an integer $m$ in such a way that
electrons are moved from the left to the right by tunneling processes with $%
t_{\mathrm{LC}}$ and $t_{\mathrm{CR}}$ under the constraint $n_{\mathrm{C}%
}\in \left\{ 0,1\right\} $. The total energy of state $\mathbf{n}=\left( n_{%
\mathrm{L}},n_{\mathrm{C}},n_{\mathrm{R}}\right) $ in the absence of
tunneling can be written as%
\begin{eqnarray}
E_{\mathbf{n}} &=&\sum_{i=\mathrm{L,C,R}}\left[ n_{i}\varepsilon _{i}+\frac{1%
}{2}n_{i}\left( n_{i}-1\right) U_{i}\right]  \label{Etot} \\
&&+n_{\mathrm{L}}n_{\mathrm{C}}U_{\mathrm{LC}}+n_{\mathrm{C}}n_{\mathrm{R}%
}U_{\mathrm{CR}}.  \notag
\end{eqnarray}%
Therefore, the matrix form of the Hamiltonian with charge bases has diagonal
elements of $E_{\mathbf{n}}$ and nearest-neighbor off-diagonal elements $t_{%
\mathrm{LC}}$ and $t_{\mathrm{CR}}$. $H_{1}$ and $H_{2}$ in Eqs. (\ref{Eq001}%
) and (\ref{Eq002}) are examples of $n=1$ and $m=1$ in the reduced Hilbert
space (only for three charge states). The conditions for identical
Hamiltonians ($H_{n}=H_{n+1}$) are $t_{\mathrm{LC}}=t_{\mathrm{CR}}$, $U_{%
\mathrm{LC}}=U_{\mathrm{CR}}$ ($\equiv U^{\prime }$), $U_{\mathrm{L}}=U_{%
\mathrm{R}}$ ($\equiv U$), $\varepsilon _{\mathrm{C}}=-nU^{\prime }$, and $%
\varepsilon _{\mathrm{L}}+\varepsilon _{\mathrm{R}}=-nU$. The straight
diagonal line is expected to appear if all conditions are met.

The last two conditions can be written with convenient but misleading
electrochemical potentials $\mu _{\mathrm{L}}=\varepsilon _{\mathrm{L}}+mU$, 
$\mu _{\mathrm{C}}=0$, and $\mu _{\mathrm{R}}=\varepsilon _{\mathrm{R}%
}+\left( n-m\right) U$ for adding an electron to QADs L, C, and R,
respectively, from charge state $\left( m,0,n-m\right) $ [the dashed arrows
in Fig. 8(a)], where the hybridization is not considered at all. Notice $\mu
_{\mathrm{L}}+\mu _{\mathrm{R}}=0$ under the required conditions, as shown
in the energy diagram of Fig. 8(b). Conventional sequential-tunneling
transport is not allowed for this condition ($\mu _{\mathrm{L}}\neq \mu _{%
\mathrm{C}}\neq \mu _{\mathrm{R}}$), and thus does not explain the
appearance of the diagonal line. In the presence of significant $t_{\mathrm{%
LC}}=t_{\mathrm{CR}}$, the charge states of $n$- and $\left( n+1\right) $%
-electron systems are strongly hybridized with the identical matrix form of
Hamiltonians, and thus the correct electrochemical potential of the triple
QAD is $\mu _{\mathrm{T}}=0$ on the diagonal line. Therefore, the
hybridization plays an essential role in the appearance of the diagonal line.

The appearance of the diagonal line implies that the system satisfies all
conditions. As our sample shows $U_{\mathrm{LC}}\simeq U_{\mathrm{CR}}$ and $%
U_{\mathrm{L}}\simeq U_{\mathrm{R}}$, $t_{\mathrm{LC}}=t_{\mathrm{CR}}$ and $%
\mu _{\mathrm{C}}=0$ must be satisfied within the experimental allowance.
Considering the variations of the patterns at different $B$'s in Fig. 5, the
data in Fig. 6(a) could be the special case close to the symmetric
conditions.

\subsection{Numerical simulation}

The current profiles under the symmetric and non-symmetric conditions are
calculated by using the standard master equation~\cite{FujisawaPhysicaE2011}%
. The two gate voltages, $\tilde{V}_{\mathrm{L}}$ and $\tilde{V}_{\mathrm{R}%
} $, control the electrostatic potentials of the three QADs, $\varepsilon _{%
\mathrm{L}}$, $\varepsilon _{\mathrm{C}}$, and $\varepsilon _{\mathrm{R}}$,
with the lever arm factors, $\alpha _{\mathrm{R}}=\alpha _{\mathrm{L}%
}=-0.13e $, $\alpha _{\mathrm{CL}}=\alpha _{\mathrm{CR}}=-0.06e$, and $%
\alpha _{\mathrm{LR}}=\alpha _{\mathrm{RL}}=0$, and offset energies $%
\varepsilon _{\mathrm{L0}}=\varepsilon _{\mathrm{R0}}=0$ and $\varepsilon _{%
\mathrm{C0}}$ [see Appendix A for their definitions]. The Hamiltonian $H_{n}$
is diagonalized to obtain the eigenstates. The current through the triple
QAD is calculated for small bias voltage $V_{\mathrm{S}}=$ $30\ \mathrm{\mu
eV}$ between the source and the drain at electron temperature $T$ ($k_{%
\mathrm{B}}T=10\ \mathrm{\mu eV}$). Tunneling rates to the source and the
drain are fixed at $\Gamma _{\mathrm{S}}=\Gamma _{\mathrm{D}}=$ 1 GHz. The
calculation scheme for the wide range of charge states is described in
Appendix B.

Figure 7(e) shows the calculated $G$ under the symmetric condition with $t_{%
\mathrm{LC}}=t_{\mathrm{CR}}=50\ \mathrm{\mu eV,}$ $U_{\mathrm{L}}=U_{%
\mathrm{R}}=200\ \mathrm{\mu eV}$, $U_{\mathrm{LC}}=U_{\mathrm{CR}}=75\ 
\mathrm{\mu eV}$, and $\varepsilon _{\mathrm{C0}}=0$, where identical
Hamiltonian $H_{n}=H_{n+1}$ is expected at $\tilde{V}_{\mathrm{L}}+\tilde{V}%
_{\mathrm{R}}=0$. The central diagonal line (marked by the arrows) is
reproduced at $\tilde{V}_{\mathrm{L}}+\tilde{V}_{\mathrm{R}}=0$. The
honeycomb pattern is clearly resolved near the line, and the splitting is
gradually decreasing toward the upper-right and lower-left corners. All
features are symmetric about the diagonal line (highlighted by the dot
pair). They are qualitatively the same as the experimental features
including the mirror symmetry in Fig. 7(b), which suggests that the
symmetric conditions are satisfied in the experiment.

When a small energy offset of $\varepsilon _{\mathrm{C0}}=$ 8.7 $\mu $eV is
introduced to the conditions for Fig. 7(e), the pattern is no longer mirror
symmetric about the diagonal line, as shown in Fig. 7(d). The pattern shows
the glide reflection symmetry (highlighted by the dot pair) about the dashed
line between the double diagonal line (marked by the arrows). Such glide
reflection symmetry is seen in our experimental data of Fig. 7(a).

Identical tunneling with $t_{\mathrm{LC}}=t_{\mathrm{CR}}$ is the essential
condition. Our simulation (not shown) suggests that we would not recognize
the deviation from the straight diagonal line if the asymmetry is not large (%
$\gamma \equiv \left\vert \frac{t_{\mathrm{LC}}-t_{\mathrm{CR}}}{t_{\mathrm{%
LC}}+t_{\mathrm{CR}}}\right\vert \lesssim 0.1$). When large $\gamma =0.4$ ($%
t_{\mathrm{LC}}=70\ \mathrm{\mu eV}$ and $t_{\mathrm{CR}}=30\ \mathrm{\mu eV}
$) is assumed in the simulation, the diagonal line disappears as shown in
Fig. 7(f).
While the honeycomb pattern is seen in the entire region of the figure, the splitting shows gentle variation. 
Similar pattern is seen in our
data of Fig. 7(c), whereas the parameters for Fig. 7(f) were not adjusted to the experimental data.
%Thus, though the honeycomb structures shown in Fig. 7(c), Fig. 5(a), and (h), appear at first glance to be double QAD, the change in splitting indicates that triple QAD transport is occurring.

Identical Coulomb interactions with $U_{\mathrm{LC}}=U_{\mathrm{CR}}$ and $%
U_{\mathrm{L}}=U_{\mathrm{R}}$ are important for the periodicity along the
diagonal line. Some diagonal lines are visible only for a few oscillation
periods, which may be related to small asymmetry in the Coulomb interactions.

Whereas we observed smooth tuning of honeycomb patterns with gate voltages,
we do not see systematic variation with the magnetic field.
Slight change in magnetic field can induce drastic change in the stability diagram, which might be related to uncontrollable charging of localized states.

\section{Summary}

We have proposed and demonstrated the triple QAD scheme for tunable coupling
between two separate QADs by using cotunneling through the central QAD. The
charge stability diagram of the system changes from the parallelogram
pattern for the uncoupled case to the round honeycomb pattern for the
coupled case by tuning the energy level of the central QAD. In a special
case, the charge diagram shows diagonal straight lines as a signature of symmetric parameters of the triple QAD. Systematic variation of transport
characteristics are studied by numerical calculation based on the master
equation and by experiment with unintentional QADs and a localized state.
The system can be made more tunable, if the localized state is replaced by
an intentional QAD with an independent gate. Our research has paved the way
for further studies on multiple QADs in the integer and fractional QH
states, such as a QAD array for anyon operations~\cite{AverinPhysicaE2007}.

\section*{Acknowledgement}

This work was supported by JSPS KAKENHI Grant No. JP19H05603 and JP19K14630,
and partially conducted at Nanofab in the Tokyo Institute of Technology
supported by ``Advanced Research Infrastructure for Materials and
Nanotechnology (ARIM)'' in Japan and at Materials Analysis Division, Open Facility Center in Tokyo Institute of Technology.

\appendix

\section{Lever arm factors}

The electrostatic potentials $\varepsilon _{\mathrm{L}}$, $\varepsilon _{%
\mathrm{C}}$, and $\varepsilon _{\mathrm{R}}$ can be changed by the gate
voltages $V_{\mathrm{L}}$ and $V_{\mathrm{R}}$ with linear relations,%
\begin{eqnarray}
\varepsilon _{\mathrm{L}} &=&\alpha _{\mathrm{L}}V_{\mathrm{L}}+\alpha _{%
\mathrm{LR}}V_{\mathrm{R}}+\varepsilon _{\mathrm{L0}}  \label{epsLCR} \\
\varepsilon _{\mathrm{C}} &=&\alpha _{\mathrm{CL}}V_{\mathrm{L}}+\alpha _{%
\mathrm{CR}}V_{\mathrm{R}}+\varepsilon _{\mathrm{C0}}  \notag \\
\varepsilon _{\mathrm{R}} &=&\alpha _{\mathrm{R}}V_{\mathrm{R}}+\alpha _{%
\mathrm{RL}}V_{\mathrm{L}}+\varepsilon _{\mathrm{R0}}  \notag
\end{eqnarray}%
with lever arm factors ($\alpha $'s) and offsets ($\varepsilon _{\mathrm{L0}%
} $ and so on). We estimated $\alpha _{\mathrm{R}}\simeq -0.13e$ from the
data in Fig. 3(e) and $\alpha _{\mathrm{L}}\simeq -0.15e$ from Fig. 3(f).
The CB oscillation periods of QAD L and R in Fig. 6(a) at \#T are similar
and close to the period of QAD R at \#R. Therefore, the lever arm factor $%
\alpha _{\Delta }=-0.13e$ was used to obtain $\bar{\Delta}=\frac{1}{2}\alpha
_{\Delta }\Delta \tilde{V}$ in Fig. 6(d). The small ratios $\alpha _{\mathrm{%
LR}}/\alpha _{\mathrm{L}}\simeq \alpha _{\mathrm{RL}}/\alpha _{\mathrm{R}%
}\simeq 0.04$ are estimated from the slope of the CB oscillations (the
dot-dashed lines for $\alpha _{\mathrm{LR}}/\alpha _{\mathrm{L}}$)\ in Fig.
6(a), but $\alpha _{\mathrm{LR}}$ and $\alpha _{\mathrm{RL}}$ are neglected
in the numerical calculations for simplicity.

Unfortunately, we have no direct estimates on $\alpha _{\mathrm{CL}}$ and $%
\alpha _{\mathrm{CR}}$. For example, $\alpha _{\mathrm{CL}}$ describing the
effect of the left gate on the QAD C potential should arise from the direct
capacitive coupling and indirect coupling through QAD L. Whereas the former
contribution is unknown, the latter can be estimated from $\alpha _{\mathrm{L%
}}\frac{U_{\mathrm{LC}}}{U_{\mathrm{L}}}\simeq -0.05e$. As $\alpha _{\mathrm{%
CL}}$ should be smaller than $\alpha _{\mathrm{L}}$, $\alpha _{\mathrm{CL}}$
as well as $\alpha _{\mathrm{CR}}$ should be in the range of $-0.05e\sim
-0.13e$. If available, these values provide $\eta ^{\prime }\simeq \frac{1}{2%
}\left( \alpha _{\mathrm{CL}}+\alpha _{\mathrm{CR}}\right) $ for the
determination $\varepsilon _{\mathrm{C}}$ in Fig. 6(d). This parameter range
does not contradict $\eta ^{\prime }=-0.06e$ obtained from the fitting to
the data in Fig. 6(d).
Therefore, we used $\alpha_{\rm CL}=\alpha_{\rm CR}=-0.06e$ in the numerical simulations. 

\section{Calculation of triple-QAD current}

We calculated the current through the triple QAD based on the master
equation~\cite{FujisawaPhysicaE2011}. The electrochemical potential $%
\varepsilon _{i}$ for the first excess electron only in QAD $i$ can be
controlled with excess gate voltages $\tilde{V}_{\mathrm{L}}$ and $\tilde{V}%
_{\mathrm{R}}$, in the form of Eq. (\ref{epsLCR}). Here, $\varepsilon _{%
\mathrm{C0}}$ plays an important role in the stability diagram, and we set $%
\varepsilon _{\mathrm{L0}}=\varepsilon _{\mathrm{R0}}=0$ and $\alpha _{%
\mathrm{LR}}=\alpha _{\mathrm{RL}}=0$ for simplicity. The Hamiltonian $H_{n}$
shown in Sec. IV-A is diagonalized to obtain $k$-th eigenenergy $E_{n,k}$ of 
$n$-electron system. Transport through the triple QAD can be calculated by
considering tunnel transitions to the leads. We assumed energy-independent
tunneling rates $\Gamma _{\mathrm{S}}$ between the source and QAD L and $%
\Gamma _{\mathrm{D}}$ between QAD R and the drain. A master equation for
occupation probabilities of the eigenstates is constructed under a small
bias voltage $V_{\mathrm{S}}=$ 30 $\mu $V and thermal energy $k_{\mathrm{B}%
}T=$ 10 $\mu $V in the leads. For each $\tilde{V}_{\mathrm{L}}$ and $\tilde{V%
}_{\mathrm{R}}$, a few eigenstates with energies in the range of $%
E_{n,k}\leqq E^{\mathrm{(min)}}+eV_{\mathrm{S}}+3k_{\mathrm{B}}T$ contribute to
the transport, where $E^{\mathrm{(min)}}$ is the total ground-state energy
for all possible $n$ and $k$. The eigenstates within this energy range are
considered in solving the master equation. The current was calculated from
the steady-state occupation probabilities.

\end{document}